\documentclass[aps,floatfix,preprint,preprintnumbers,nofootinbib,superscriptaddress,natbib]{revtex4}

\usepackage[titletoc]{appendix}
\usepackage{graphicx,float}
\usepackage{epsfig}		
\usepackage{dcolumn}
\usepackage{bm}
\usepackage{subfigure}
\usepackage[all]{xy}
\usepackage{amsmath,upgreek}
\usepackage{amssymb}
\usepackage{pdfpages}
\usepackage{color}
\usepackage{graphicx,epstopdf}
\usepackage[colorlinks=true,
 linkcolor=red,
 urlcolor=purple,
 citecolor=blue,hyperindex]{hyperref}

\def\0{\mbox{\tiny $0$}}
\def\1{\mbox{\tiny $1$}}
\def\2{\mbox{\tiny $2$}}
\def\3{\mbox{\tiny $3$}}
\def\4{\mbox{\tiny $4$}}
\def\5{\mbox{\tiny $5$}}
\def\6{\mbox{\tiny $6$}}
\def\7{\mbox{\tiny $7$}}
\def\8{\mbox{\tiny $8$}}
\def\9{\mbox{\tiny $9$}}

\def\f14{\mbox{\tiny $\frac{1}{4}$}}

\begin{document}
\title{Hubble tension problem encompassed by phase-space quantum cosmology} 
\author{Alex E. Bernardini}
\email{alexeb@ufscar.br}
\affiliation{~Departamento de F\'{\i}sica, Universidade Federal de S\~ao Carlos, PO Box 676, 13565-905, S\~ao Carlos, SP, Brasil.}
\altaffiliation[Also at]{~Departamento de F\'isica e Astronomia, Faculdade de Ci\^{e}ncias da Universidade do Porto, Rua do Campo Alegre 687, 4169-007, Porto, Portugal.}
\date{\today}
 
\begin{abstract}
Analytical solutions encompassing the so-called Hubble tension problem are revisited through the framework of Weyl--Wigner quantum mechanics and discussed in the context of generalized phase-space scenarios of quantum cosmology.
After reviewing the nature of the problem and its recent developments, an extended formulation constructed within the quantum phase-space framework to address the Hubble tension is proposed.
For the quantum cosmology described in the minisuperspace framework through (generic) localized phase-space quantum states, when residual quantum corrections to the Einstein--Friedmann equation are analytically derived, quantum effects are shown to suppress the Hubble tension divergence between early- and late-time predictions.
Besides addressing the Hubble tension problem within the standard $\Lambda$CDM cosmological model, our approach encompasses generalized quantum cosmological scenarios that also include curvature and dark sector modifications.
\end{abstract}

\date{\today}
\maketitle

\section{Introduction}

The so-called {\em Hubble tension} refers to the discrepancy between the value of the Hubble constant, $H_0$, inferred from the Planck collaboration's observations of the cosmic microwave background (CMB), interpreted within the $\Lambda$CDM model of the early Universe, which predicts $H_0 = (67.27 \pm 0.60)\, \mathrm{km/(s\cdot Mpc)}$ \cite{Aghanim}, and the value obtained by the SH0ES collaboration from the Cepheid-calibrated cosmic distance ladder, $H_0 = (73.2\pm 1.3)\, \mathrm{km/(s\cdot Mpc)}$ \cite{Riess:2020fzl}. In this context, {\em early-Universe calibrations} combining measurements of Big Bang nucleosynthesis (BBN) with baryonic acoustic oscillation (BAO) data \cite{2019JCAP,2013MNRAS.436.1674A,2015PhRvD..92l3516A,Blomqvist:2019rah,2019JCAP...10..044C}, or with supernova constraints \cite{2017MNRAS.467..731V,2021PhRvD.103j3533B}, consistently predict $H_0$ values below $70\, \mathrm{km/(s\cdot Mpc)}$. Conversely, direct measurements of the local expansion rate, which avoid potential biases associated with Cepheid observations \cite{Lesgougues,Freedman2019jwv,Huang2019yhh,Khetan2020hmh,Hubble1,Hubble2,Hubble3,Aghanim,Alam2021,Abbott2022,7777,8888,9999,pheno}, generally yield $H_0$ values much closer to those reported by SH0ES. In particular, observational programs that systematically favor $H_0$ values higher than the Planck prediction do not completely exclude the possibility that the discrepancy originates from yet unidentified systematic effects \cite{Hubble1,Hubble2,Hubble3,Vagnozzi}. However, this interpretation has progressively lost support following the validation by the James Webb Space Telescope (JWST) of Hubble Space Telescope distance measurements for supernova subsamples, as reported in Ref.~\cite{Riess24}. 

Since analyses of systematic effects remain insufficient for fully addressing the problem, a theoretical re-examination of the $\Lambda$CDM model operating in either the early- or late-time Universe has been considered.
Our perspective is that solutions from phase-space quantum cosmology can be admitted \cite{BerBer,Ber25} for discussing the Hubble tension.
In this paper, analytical solutions to the Hubble tension problem are revisited within the framework of Weyl--Wigner quantum mechanics and extended to phase-space scenarios of quantum cosmology which include curvature contributions.
Within a formulation validated for localized phase-space quantum states, residual quantum corrections to the Einstein--Friedmann equation are analytically evinced.
By considering corrections of quantum origin constructed upon the Weyl--Wigner phase-space framework, predictions for $H_0$ suggest a smooth interpolation between the phenomenological values obtained from early- and late-time observations, thereby alleviating the Hubble tension.
When associated with curvature and dark sector modifications, it can be shown that quantum effects attenuate the divergence between early- and late-time predictions by reducing the discrepancies as observed within the $\Lambda$CDM cosmology.

The letter is organized as follows. Section II reviews the nature of the Hubble tension from the theoretical perspective of two distinct approaches aimed at reconciling early- and late-time determinations of $H_0$. Section III summarizes the relevant aspects of Weyl--Wigner quantum mechanics and its extension to phase-space quantum cosmology. Quantum corrections to standard cosmological models, including curvature contributions, are discussed in Section IV. Finally, concluding remarks are presented in Section V, where issues related to dark-sector modifications are also addressed.
\section{The Hubble tension from the theoretical perspective}

Considering a peaked adiabatic matter overdensity at an arbitrary point in an initially homogeneous Universe, the associated baryon--photon fluid generates a shock wave propagating at the sound speed $c_s = dp/d\rho$, where $\rho$ and $p$ are the energy density and pressure. At baryon--photon decoupling (last scattering, LS), $t_{_{\rm LS}} \simeq 4\times10^{5}$ years after the Big Bang, the shock-induced overdensity reaches a radius $\sim c_s t_{_{\rm LS}}$.

The solution of the fluid equations for a peaked initial perturbation yields the Green's function governing primordial fluctuations, whose evolution produces the oscillatory features known as baryon acoustic oscillations (BAO), observed as acoustic peaks in the CMB angular power spectrum $C_l$. Measurements of these peaks allows the inference of $H_0$ through comparisons between predicted and observed $C_l$. Hence, understanding how $H_0$ is extracted from CMB data is essential for describing the Hubble tension.

A central scale in $\Lambda$CDM is the angular size of the sound horizon at recombination, $\theta_s=(1.04109\pm0.00030)\times10^{-2}$, the most precisely measured CMB parameter. The multipole of the first acoustic peak, $\ell_s$, determines the angle subtended by the sound horizon at the LS surface through $\ell_s \simeq 2/\theta_s$. This angle is given by $\theta_s=r_s/D_A$, where $D_A$ is the angular-diameter distance and $r_s\sim c_s t_{_{\rm LS}}$ is the comoving sound horizon:
\begin{equation}\label{eq:soundhorizon}
 r_s = \int_{z_{_{\rm LS}}}^\infty \frac{c_s(z)\, dz}{H(z)} = \frac{c}{\sqrt{3}H_{_{\rm LS}}} \int_{z_{_{\rm LS}}}^\infty {dz} {\left[\frac{\rho(z)}{\rho(z_{_{\rm LS}})} \left(1 + \frac{3\omega_b}{4\omega_\gamma}\frac{1}{1+z}\right) \right]^{-1/2}},
\end{equation}
with $c_s(z) = c \left[ 3\left(1+\frac{3\omega_b}{4\omega_\gamma}\frac{1}{1+z}\right) \right]^{-1/2}$, where $\rho(z)$ is the total energy density at a redshift $z$, $\omega_b= \Omega_b h^2$ is the physical baryon density today, $\omega_b=0.0224\pm0.0001$, determined by the higher-peak structure in the CMB power spectrum, and $\omega_\gamma =2.47 \times 10^{-5}$ is the physical photon energy density from Planck's $\Lambda$CDM.
The expansion rate at the CMB photons LS redshift ($z_{_{\rm LS}}\simeq 1080$) is $H_{_{\rm LS}} = H_0 \,h^{-1} (1+z_{_{\rm LS}})^2 \sqrt{ \omega_r +\frac{\omega_m}{1+z_{_{\rm LS}}} }$, where $h\equiv H_0/(100\, {\rm km}\, {\rm sec}^{-1}\, {\rm Mpc}^{-1})$ is the dimensionless form for the Hubble constant, $\omega_m=\Omega_m h^2$ is the physical nonrelativistic-matter density today, $\omega_m = 0.142\pm 0.001$ (also fixed fairly precisely by the higher-peak structure in the CMB\footnote{$\omega_m$ and $\omega_b$ are determined primarily by characteristics in the CMB power spectrum such as the Silk damping at higher $l$ and the relative heights of the even- and odd-numbered peaks.}), and 
$\omega_r$ is the physical radiation density, $\omega_r = \omega_\gamma\left[ 1 + \frac78 N_{\rm eff} \left( \frac{4}{11} \right)^{4/3} \right]$ (with $N_{\rm eff}\sim3.06$ accounting for neutrino mass eigenstate contributions \cite{Dodelson}). Such contributions summarize the early-Universe energy density as $\rho(z) \propto \omega_m (1+z)^3 + \omega_r(1+z)^4$, for which tiny dark energy contributions ($z > z_{_{\rm LS}}$)\footnote{Small corrections on $z_{_{\rm LS}}$ due to contributions from $\omega_b$ and $\omega_m$ are not relevant for the Hubble tension.} are ignored. 
The complementary cosmological timeline is depicted by the angular-diameter distance in the comoving form
\begin{equation}
 D_A = \frac{c}{H_0} \int_{0}^{z_{_{\rm LS}}}
 \frac{dz}{ \left[\rho(z)/\rho_0 \right]^{1/2}},
\label{angulardiameterdistance}
\end{equation}
where the total energy density, $\rho(z)$, which is now relevant, runs from the period from recombination until the present time, $z=0$, such that radiation contributions can be suppressed as to have $\rho(z)/\rho_0 \sim \Omega_m(1+z)^3+(1-\Omega_m)(1+z)^{-3(1+w)}$, where a dark-energy equation-of-state parameter, $w$, has been introduced\footnote{The cosmological constant corresponds to $w=-1$.}.

Finally, from $\theta_s = r_s/D_A$, one infers the Hubble constant at late times (as from Eq.~\eqref{angulardiameterdistance}),
\begin{eqnarray}
 H_0 &=& \sqrt{3} H_{_{\rm LS}} \theta_s \frac{ \int_{0}^{z_{_{\rm LS}}}
 dz\, \left[\rho(z)/\rho_0 \right]^{-1/2} }
 { \int_{z_{_{\rm LS}}}^\infty dz\, \left[\rho(z)/\rho(z_{_{\rm LS}}) \right]^{-1/2} \left[1 + {3\omega_b}/({4\omega_\gamma}{(1+z)})\right]^{-1/2} },
 \label{cmbH0}
\end{eqnarray}
from which, given that $h \equiv h_{_{LT}}=H_0/(100 {\rm km}~{\rm sec}^{-1}~{\rm Mpc}^{-1})$, an implicit form for determining $h$ from $H_{_{LS}}$, $h\equiv h_{_{LS}}$, is written as
 \begin{eqnarray}
 h_{_{LS}} 
&=& \frac{ \int_{z_{_{\rm LS}}}^\infty {dz} {\left[\left(\omega_m (1+z)^3 + \omega_r(1+z)^4\right) \left(1 + {3\omega_b}/({4\omega_\gamma}{(1+z)})\right) \right]^{-1/2}} }{\sqrt{3} \theta_s { \int_{0}^{z_{_{\rm LS}}}
 dz\, \left[\omega_m(1+z)^3+(h_{_{LT}}^2-\omega_m)(1+z)^{-3(1+w)} \right]^{-1/2}}}h_{_{LT}}\label{cmbh}
\end{eqnarray}
where Eq.~\eqref{cmbH0} has been used and $\rho_\mathrm{Crit} \equiv \rho_{0}$.

Ref.~\cite{Lesgougues} discusses two different approaches for understanding how both data sets, from CMB and BAO, constrains $h(H_0)$.
In the first and less restrictive approach, physical energy densities contribute to the Hubble parameter $H(z)$ through the Friedmann equation.
Considering that the energy density today is dominated by dark energy, it is argued that a change in $H_0$ is completely constrained by a re-scaling of $\Omega_\mathrm{\Lambda}$ within a flat-$\Lambda$CDM model \cite{Lesgougues}. 
Since the sound horizon of \eqref{eq:soundhorizon} is not impacted by $H_0$ in this approach, increasing the value of $h(H_0)$ via a late-time modification of dark energy would impose that the energy density should be smaller in the past in order to constrain the angular diameter distance which should be fixed as to not affect the measured angular scales \cite{Lesgougues}. The problem, in this case, would emerge from the multitude of the low redshift measurements which tightly constrain $H(z)/H_0$ in a redshift range of around $z \lesssim 2$. Constraining the angular diameter distances to fixed values at redshifts from $\sim 1$ to $\sim 1000$ varying $H_0$ in this approach is an exceptionally difficult task.
In the second approach \cite{Lesgougues}, given that most low redshift probes just constrain $H(z)/H_0$, and not (either) $H$ (or $H_0$) as expected from Eq.~\eqref{cmbh}, the computation of $H_0$ is just sensitive to the fractional densities $\Omega_X=\rho_X/\rho_\mathrm{Crit}$ rather than the physical ones, $\rho_X$. Hence, satisfying the first approach implies in satisfying the second one. From the theoretical perspective here considered, this is the fiducial point of our analysis.

\section{Phase-space quantum mechanics and the effective quantum potential}

Supported by the Heisenberg-Weyl algebra (cf. the position-momentum non-commutative relation, $[x,\,k] = i$), the Wigner {\em quasi}-distribution function, $\mathcal{W}(k,\, p)$, is written in terms of dimensionless canonical coordinates of position, $x$, and momentum, $k$, through the Weyl transform of the quantum density matrix operator, $\hat{\varrho} = |\psi \rangle \langle \psi |$, such that \cite{Wigner,Ballentine,Case}
\begin{equation}
\hat{\varrho} \to \mathcal{W}(x,\, k) = \pi^{-1} 
\int^{+\infty}_{-\infty} \hspace{-.35cm}dq\,\exp{\left[2\, i \, k \,q\right]}\,
\psi(x - q)\,\psi^{\ast}(x + q).\label{222}
\end{equation}

Relevant to the following analysis, the Wigner function satisfies the continuity equation\cite{Case,Ballentine,Steuernagel3,NossoPaper,Meu2018},
\begin{equation}\label{z51dim}
{\partial_{\tau} \mathcal{W}} + {\partial_x \mathcal{J}_x}+{\partial_k \mathcal{J}_k} = {\partial_{\tau} \mathcal{W}} + \mbox{\boldmath $\nabla$}_{\xi}\cdot\mbox{\boldmath $\mathcal{J}$} =0,
\end{equation}
where the time variable, $\tau$, is also dimensionless. For generalized Hamiltonians cast as $\mathcal{H}(x,\,k) = \mathcal{K}(k) + \mathcal{V}(x)$, 
the corresponding Wigner currents are given by \cite{Novo2021B}
\begin{eqnarray}
\label{imWAmm}\mathcal{J}_x(x, \, k;\,\tau) &=& +\sum_{\eta=0}^{\infty} \left(\frac{i}{2}\right)^{2\eta}\frac{1}{(2\eta+1)!} \, \left[\partial_k^{2\eta+1}\mathcal{K}(k)\right]\,\partial_x^{2\eta}\mathcal{W}(x, \, k;\,\tau),\\
\label{imWBmm}\mathcal{J}_k(x, \, k;\,\tau) &=& -\sum_{\eta=0}^{\infty} \left(\frac{i}{2}\right)^{2\eta}\frac{1}{(2\eta+1)!} \, \left[\partial_x^{2\eta+1}\mathcal{V}(x)\right]\,\partial_k^{2\eta}\mathcal{W}(x, \, k;\,\tau),
\end{eqnarray}
from which the quantum back reaction is quantified by the $\eta \geq 1$ contributions from the corresponding series expansions \cite{Steuernagel3,NossoPaper,Meu2018}.
The quantum back reaction resumed by Eqs.~\eqref{imWAmm} and \eqref{imWBmm}, when both correspond to convergent series expansions, can be read as an effective quantum modification to the classical potential, $\mathcal{V}(x)$. In a subsequent effective quantum analysis, $\mathcal{V}(x)$ is then replaced by $\mathcal{U}(x)$, which includes all the non-linear contributions, $\left[\partial_x^{2\eta+1}\mathcal{V}(x)\right]$, as described by Eq.~\eqref{imWBmm}. In this case, one would have
${v}_{k(\mathcal{C})} = \dot{k}\equiv -{\partial_x \mathcal{H}}=-{\partial_x \mathcal{V}}$ being replaced by
\begin{eqnarray}
\label{imWBmmw} w_k(x, \, k;\,\tau) = -\partial_x \mathcal{U}&=& -\sum_{\eta=0}^{\infty} \left(\frac{i}{2}\right)^{2\eta}\frac{1}{(2\eta+1)!} \, \left[\partial_x^{2\eta+1}\mathcal{V}(x)\right]\,\frac{\partial_k^{2\eta}\mathcal{W}(x, \, k;\,\tau)}{\mathcal{W}(x, \, k;\,\tau)},
\end{eqnarray}
so that all the quantum corrections are encompassed by the above implicit definition of $\mathcal{U}(x)$.
Of course, for $\mathcal{K}(k) = k^2$, the corresponding series Eq.~\eqref{imWAmm} is truncated at $\eta = 0$, without affecting the analysis \cite{Novo2021B}.
Nevertheless, the manipulation of Eq.~\eqref{imWBmmw} works fine only for $\partial_k^{2\eta}\mathcal{W}(x, \, k;\,\tau)/\mathcal{W}(x, \, k;\,\tau)$ matching the conditions which leads to $\mathcal{U} \equiv \mathcal{U}(x)$, i.e. independent of the momentum coordinate $k$.
This can only be achieved for Wigner functions, $\mathcal{W}(x, \, k;\,\tau)$, identified as sinusoidal and hyperbolic functions of the product $k\,x$, multiplied by some arbitrary $x-$ and $\tau$-dependent function, $g(x;\,\tau)$\footnote{The purpose of the adopted Wigner profiles is not to reconstruct a specific microscopic wave function of the Universe. Rather, they provide analytically controllable realizations of smeared phase-space quantum states whose residual interference structure induces soft quantum back-reaction corrections to the effective cosmological energy density. In particular, oscillatory Wigner structures naturally arise in coherent superpositions, squeezed states, and semiclassical Wheeler-DeWitt (WDW) wave packets, where interference effects generate characteristic sinusoidal modulations in phase space. The hyperbolic configurations provide a complementary class of nonoscillatory deformations, allowing the investigation of qualitatively different phase-space geometries.}.

\subsection{Quantum corrections for sinusoidal Wigner distributions}

By replacing the Wigner distribution at Eq.~\eqref{imWBmmw} by a sinusoidal one, $\mathcal{W}(x, \, k;\,\tau) \equiv g(x;\,\tau)\, \mbox{Sn}(\mu \,k\,x)$, where $\mu$ is a free parameter which modulates an intrinsic phase-space coherence scale, and $\mbox{Sn}(\dots)$ is identified as either $\sin(\dots)$ or $\cos(\dots)$, after some straightforward math manipulations, one obtains \cite{Ber25}
\begin{eqnarray}\label{imWBmmws1}
\partial_x\mathcal{U}(x)&=& \sum_{\eta=0}^{\infty} \left(\frac{\mu\,x}{2}\right)^{2\eta}\frac{1}{(2\eta+1)!} \, \left[\partial_x^{2\eta+1}\mathcal{V}(x)\right].
\end{eqnarray}
For the classical potential $\mathcal{V}(x)$ written as a sum of polynomial and inverse polynomial contributions, $\mathcal{V}(x) = \sum_{\kappa} a_\kappa x^{-\kappa}$, with $\kappa \in \mathbb{Z}$, additional simplifications are evinced in order to give
\begin{eqnarray}\label{imWBmmws2}
\partial_x\mathcal{U}(x)
&=& - \sum_{\kappa} \left\{a_\kappa \sum_{\eta=0}^{\infty} \left(\frac{\mu}{2}\right)^{2\eta}\frac{\Gamma(2\eta+\kappa+1)}{\Gamma(\kappa)\Gamma(2\eta+2)}\right\} \,x^{-(\kappa+1)},
\end{eqnarray}
which, once integrated in $x$ \cite{Ber25}, 
results in the form of
\begin{eqnarray}\label{imWBmmws4}
\mathcal{U}(x)&=& \sum_{\kappa} b_\kappa(\mu) x^{-\kappa},
\end{eqnarray}
with \cite{Ber25}
\begin{eqnarray}\label{imWBmmws5}
b_\kappa(\mu) &=& a_\kappa \sum_{\eta=0}^{\infty} \left(\frac{\mu}{2}\right)^{2\eta}\frac{\Gamma(2\eta+\kappa+1)}{\Gamma(\kappa+1)\Gamma(2\eta+2)} = \frac{2^\kappa a_\kappa}{\mu\,\kappa}\left[(2-\mu)^{-\kappa}-(2+\mu)^{-\kappa}\right], \end{eqnarray}
which constrains $\mu$ to $\vert\mu\vert \in[0,\,2)$ in order to not promote drastic topological changes to the quantum potential as, of course, it should be expected from the convergence criterium for the above series expansion (with $b_\kappa(\mu) = b_\kappa(-\mu)$, which also exhibits an analytical continuation for $\mu = 0$ (with $\kappa \in \mathbb{R}$, in this case).

\subsection{Quantum corrections for hyperbolic Wigner distributions}

Similarly, by replacing the Wigner distribution at Eq.~\eqref{imWBmmw} by a hyperbolic one, $\mathcal{W}(x, \, k;\,\tau) \equiv g(x;\,\tau)\, \mbox{Snh}(\mu \,k\,x)$, where $\mbox{Snh}(\dots)$ is identified as either $\sinh(\dots)$ or $\cosh(\dots)$, one obtains
\begin{eqnarray}\label{imWBmmwh1}
\partial_x\mathcal{U}_{hyp}(x)&=& \sum_{\eta=0}^{\infty} \left(\frac{i\,\mu\,x}{2}\right)^{2\eta}\frac{1}{(2\eta)!} \, \left[\partial_x^{2\eta+1}\mathcal{V}(x)\right].
\end{eqnarray}
Again, with $\mathcal{V}(x) = \sum_{\kappa} a_\kappa x^{-\kappa}$, and $\kappa \in \mathbb{Z}$, a corresponding effective quantum potential, $\mathcal{U}_{hyp}(x)$, from quantum corrections for hyperbolic (hyp) Wigner distributions, can be obtained from $\mathcal{U}(x)$ buy turning $\mu$ into $i \mu$ in Eq.~\eqref{imWBmmws5}, such that
\begin{eqnarray}\label{imWBmmwh4}
\mathcal{U}_{hyp}(x)&=& \sum_{\kappa} c_\kappa(\mu) x^{-\kappa},
\end{eqnarray}
with
\begin{eqnarray}\label{imWBmmwh5}
c_\kappa(\mu) &=& \frac{2^\kappa}{i\,\mu\,\kappa}\left[(2-i\,\mu)^{-\kappa}-(2+i\,\mu)^{-\kappa}\right] a_\kappa = \frac{2 a_\kappa}{\mu\,\kappa({1+\mu^2/4})^{\kappa/2}}\sin\left[\kappa\arctan(\mu/2)\right],
\end{eqnarray}
without additional mathematical (not necessarily physical) constraints to the $\mu$ values (cf. arguments in Ref. \cite{Ber25}).

\subsection{Quantum corrections for standard model cosmologies and the Hubble tension solution}

The quantum mechanical problem in the context of the WDW framework \cite{DeWitt67,Hartle83,Linde84} can be introduced, for instance, from a minisuperspace action cast in the form of
\begin{equation}
S_{SM}=\frac{1}{2}\int dt \left(\frac{N}{a}\right)\left[-\left(\frac{a}{N}\dot{a}\right)^{2} +{\Omega_{_{K}}}a^{2} -\Omega_{_{\Lambda}}a^{4}-\Omega_{_{r}}-{\Omega_{_{m}}}{a} \right],
\label{eqn10SM}
\end{equation}
from which one notices that ${\Omega_{_{K}}} > 0$ stands for the curvature coupling constant and the sign of $\Omega_{_{\Lambda}}$ follows the sign of the cosmological constant. 
From $S_{SM}$, one identifies the canonical conjugate momentum associated to $a$ as given by
\begin{equation}
\Pi_{a}=\frac{\partial \mathcal{L}}{\partial \dot{a}}= - \frac{a}{N}\dot{a},
\label{eqn11SM}
\end{equation}
such that the minisuperspace Hamiltonian density becomes
\begin{equation}
 \mbox{H}=\Pi_{a}\dot{a}-\mathcal{L}=\frac{1}{2}\frac{N}{a}\left(-\Pi_{a}^{2}-{\Omega_{_{K}}}a^{2} +\Omega_{_{\Lambda}}a^{4}+\Omega_{_{r}}+{\Omega_{_{m}}}{a^{2}}\right).
\label{eqn12SM}
\end{equation}
Through the canonical quantization strategy \cite{DeWitt67,Hartle83}, the momentum $\Pi_{a}$ is promoted to an operator \cite{Hartle83}, 
$\Pi_{a}\mapsto -i \frac{d}{d a}$ such that $\quad\Pi_{a}^{2}=-\frac{1}{a^{q}}\frac{d}{d a}\left(a^{q}\frac{d}{d a}\right)$,
where the choice of $q$ does not affect the semiclassical analysis \cite{KolbTurner:1989}.
The classical minisuperspace Hamiltonian thus acts on the wave function of the Universe $\tilde{\psi}(a)$ such that the WDW equation is then given by $\mbox{H}\tilde{\psi}(a)=0$, from which one identifies the potential given by 
\begin{equation}
V(a)=\frac{1}{2}\left({\Omega_{_{K}}}a^{2}-\Omega_{_{\Lambda}}a^{4}-\Omega_{_{r}}-{\Omega_{_{m}}}{a}\right),
\label{eqn15SM}
\end{equation}
such that the classical potential, $\mathcal{V}(x)$, is related to $V(a)$ by $ \mathcal{V}(x) = - x^{-\sigma} V^{({\Omega_{_{K}}}\to 0)}(a)$ (with $x \equiv a$), being identified by
\begin{eqnarray}
\mathcal{V}(x) = x^{(4-\sigma)}\rho(x)/\rho_\mathrm{Crit} 
&=&x^{-\sigma}h^{-2}\left((h^2-\omega_{_{r}}-\omega_{_{m}})\, x^{4}+\omega_{_{r}}\,+{\omega_{_{m}}}\,{x}\right),
\label{eqn17SM}
\end{eqnarray}
with $\sigma$ constrained by the choice of the lapse function, $N$, and where the curvature contribution shall be treated separately, given that it is not relevant to our point for now. 
From Eqs.~\eqref{imWBmmws4} and \eqref{imWBmmwh4}, one would have sinusoidal and hyperbolic (associated) effective quantum potentials,
\begin{eqnarray}\label{imWBmmws4QC}
\mathcal{U}(x)
&=& x^{-\sigma}h^{-2}\left((h^2-\omega_{_{r}}-\omega_{_{m}})\,b_{\sigma-4}(\mu)\, x^{4}+\omega_{_{r}}\,b_{\sigma}(\mu)\,+{\omega_{_{m}}}\,b_{\sigma-1}(\mu)\,{x}\right),\\
\label{imWBmmwh4QC}
\mathcal{U}_{hyp}(x)
&=&x^{-\sigma}h^{-2}\left((h^2-\omega_{_{r}}-\omega_{_{m}})\,c_{\sigma-4}(\mu)\, x^{4}+\omega_{_{r}}\,c_{\sigma}(\mu)\,+{\omega_{_{m}}}\,c_{\sigma-1}(\mu)\,{x}\right),\end{eqnarray}
respectively. The coefficients $b_\kappa(\mu)$ and $c_\kappa(\mu)$, are given by Eqs.~\eqref{imWBmmws5} and \eqref{imWBmmwh5}, respectively. According to our hypothesis, $b_\kappa(\mu)$ and $c_\kappa(\mu)$ are one parameter functions of $\mu$, which modulate the modifications on the cosmic energy density pattern, for expected fixed values of the Hubble parameter $h$.
The coefficients $c_{\sigma-4}(\mu)$ and $b_{\sigma-4}(\mu)$ are indeed set equal to unity, since the smooth quantum effects are assumed to be suppressed at very late times ($x \lesssim 1$). As a net effect, one could read the results from quantum corrections, Eqs.~\eqref{imWBmmws4QC} and \eqref{imWBmmwh4QC},
as $\mathcal{U}(x)= (\tilde{h}/h_{_{LT}})^{-2} \mathcal{V}(x)$ and $\mathcal{U}_{hyp}(x)= (\tilde{h}/h_{_{LT}})^{-2} \mathcal{V}(x)$, respectively (in both cases, with $\equiv \tilde{\rho}(x)= (\tilde{h}/h_{_{LT}})^{-2}\rho(x)$), which would be misinterpreted as a modulation from a modified Hubble parameter, $\tilde{h}$ (cf. Fig.~01 in Ref.~\cite{Ber25}). 
 
\begin{figure}[h!]
\includegraphics[scale=.23]{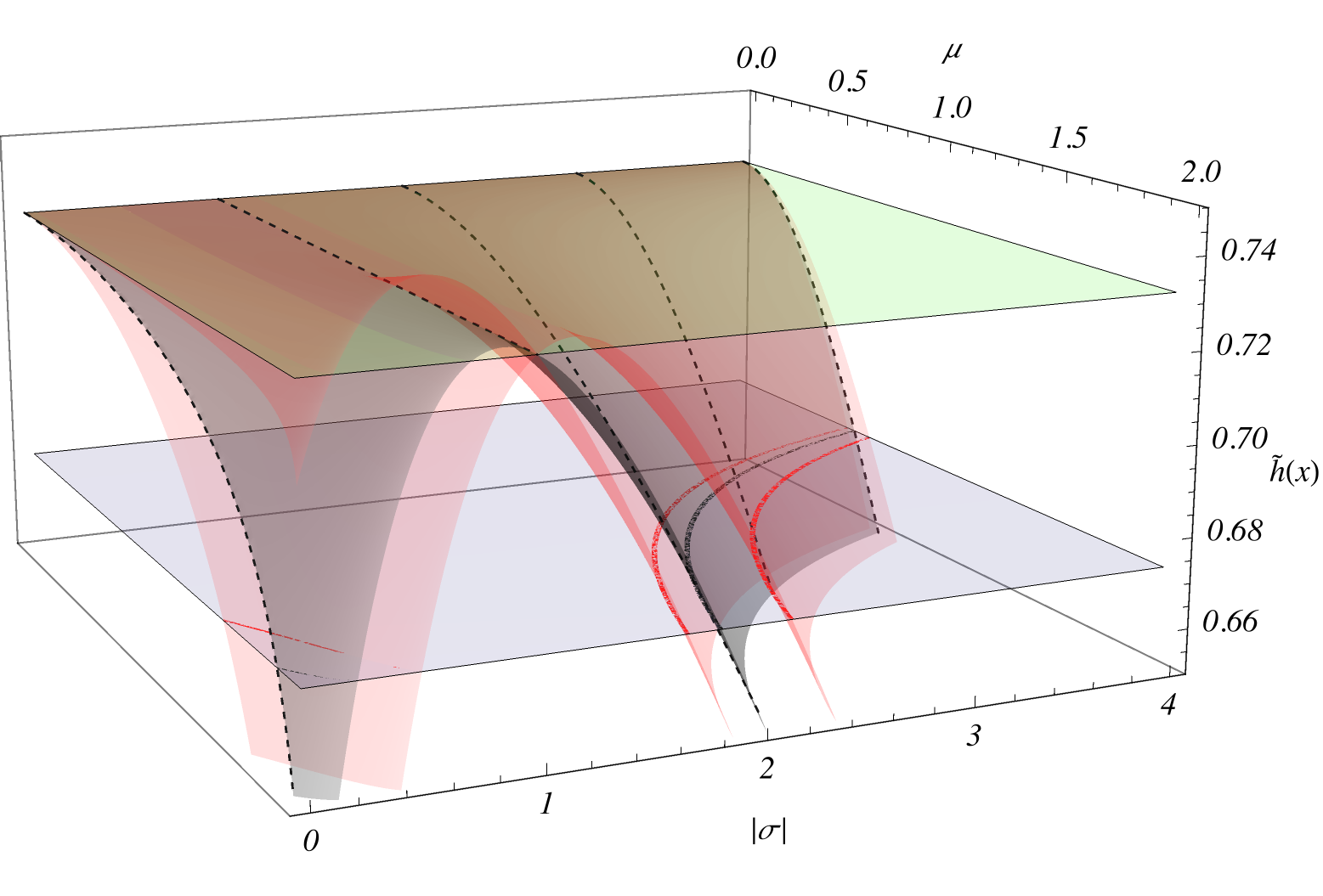}
\includegraphics[scale=.23]{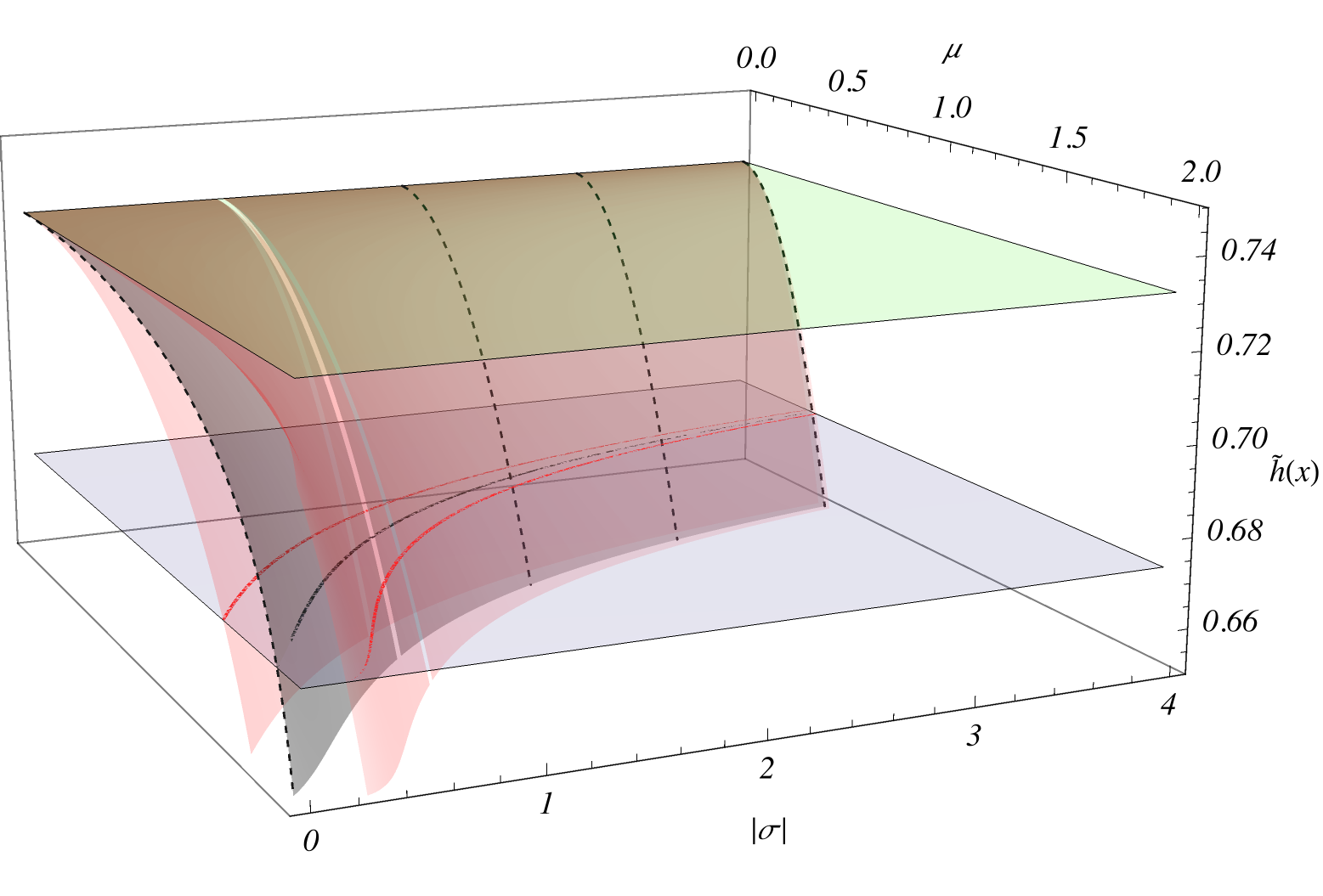}
\includegraphics[scale=.45]{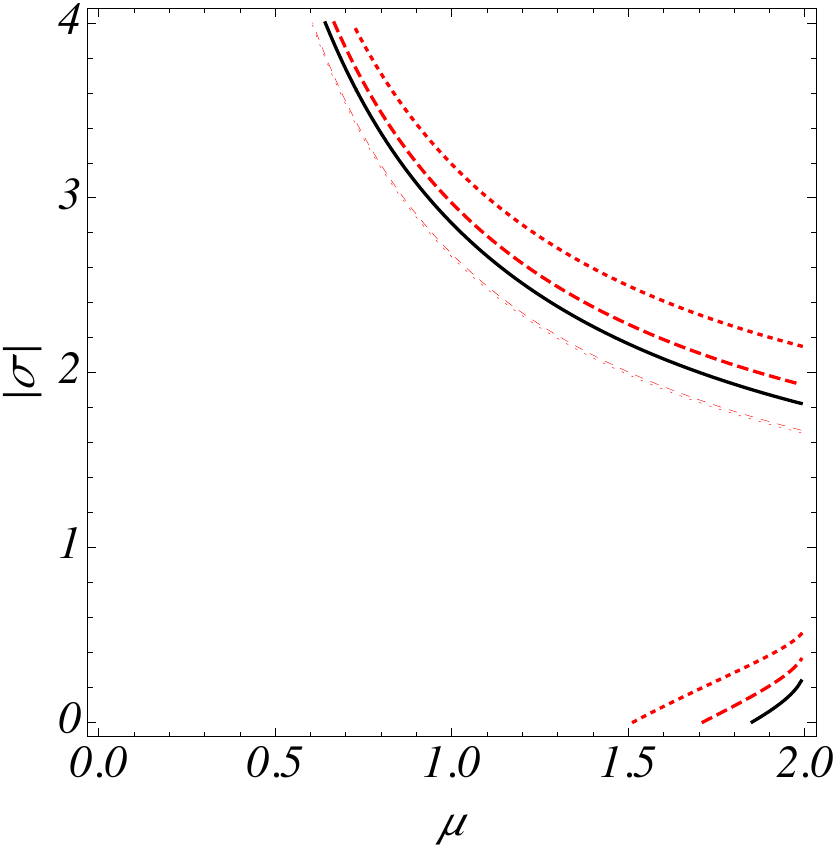}
\includegraphics[scale=.45]{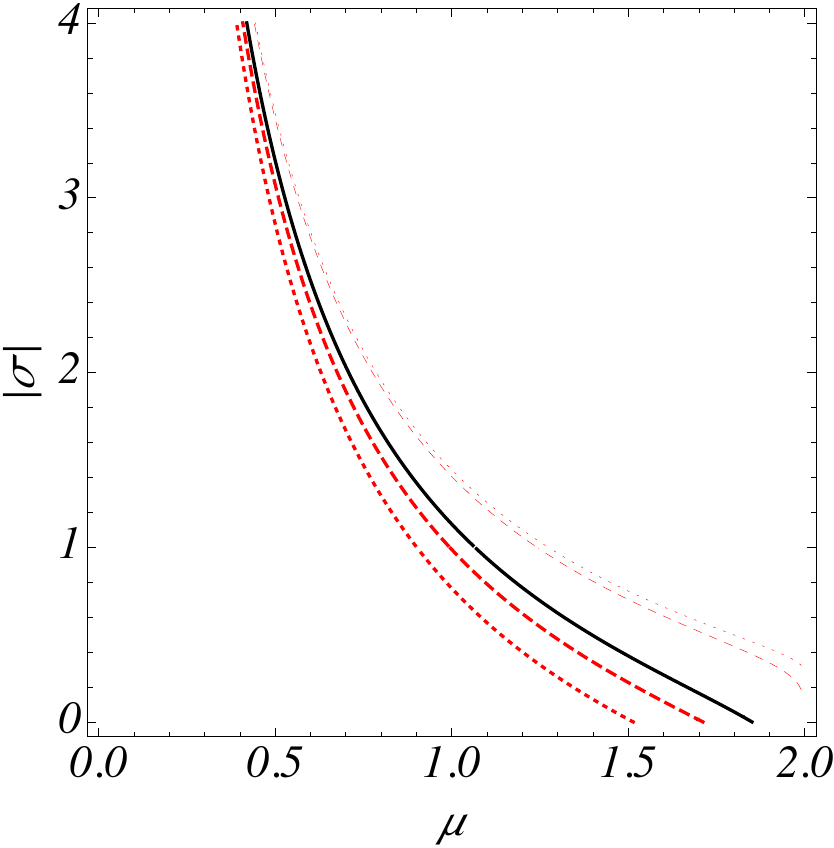}
\renewcommand{\baselinestretch}{1}
\caption{\footnotesize(Color online) (First row) $\sigma$ and $\mu$ constraints such that $\tilde{h}(x\sim1) \to h_{_{LT}} = 0.732$ (green plane) and $\tilde{h}(10^{-4} \lesssim x \lesssim 10^{-1}) \sim 0.673$ (blue plane) for $\sigma < 0$ (first column) and $\sigma > 0$ (second column). Dashed black lines denote integer values for $\vert \sigma \vert$. 3D surfaces show $\tilde{h}(x)$ as function of $\sigma$ and $\mu$ for $x = (1+1080)^{-1}$ (black surface) and $x = 10^{-2}$ and $x = 2\times 10^{-4}$ (red surfaces).
(Second row) Results correspond to the projection on the $\tilde{h}((1+1080)^{-1}) \sim 0.673$ plane (black thick line), i.e. the LS surface at $z = 1080$. Red dashed and dotted lines depict the smooth changes of $\tilde{h}(x)$ as function of $x$. In this case, $\sigma$ and $\mu$ constraints are for $\tilde{h}(10^{-1})$ (dotted thin), $\tilde{h}(10^{-2})$ (dashed thin), $\tilde{h}(5 \times 10^{-4})$ (dashed thick) and $\tilde{h}(2 \times 10^{-4})$ (dotted thick).
}\label{terca}
\end{figure}

Results from $\mathcal{U}(x)$ (sinusoidal modulation) exhibit the phenomenologically expected well-defined {\em transient plateaus} for $\tilde{h}$ from $\sim 0.673$ to $\sim 0.732$ at early and late times, respectively\footnote{Otherwise, results from $\mathcal{U}_{hyp}(x)$ (hyperbolic modulation) predict an opposite behavior from that predicted by the phenomenology, and therefore can be discarded.}. As depicted in Fig.~\ref{terca}, parameters $\sigma$ and $\mu$ can be constrained one to each other as to return the expected results for early and late times, i.e. $\tilde{h}^2(1) \sim 0.732 \pm 0.013$ and $\tilde{h}^2(x\ll 1) \sim 0.673 \pm 0.006$. The phenomenological parameters were set as $h = 0.732$, $\omega_m = 0.142\pm 0.001$ and $\omega_\gamma =2.47 \times 10^{-5}$.

In particular, $\sigma$ and $\mu$ constraints are for $\tilde{h}(10^{-1})$ (dotted thin), $\tilde{h}(10^{-2})$ (dashed thin), $\tilde{h}(5 \times 10^{-4})$ (dashed thick) and $\tilde{h}(2 \times 10^{-4})$ (dotted thick).

Even if the above analysis explains the observed divergent results for the Hubble parameter at early and late times through a one degree of freedom constraint between $\sigma$ and $\mu$ parameters, this just bring up some elementary modifications to the standard cosmological model, in particular, when it is driven by a Hamiltonian approach.
For the sinusoidal configurations, the parameter $\mu$ characterizes the intrinsic phase-space coherence scale associated with the quantum cosmological state. Within the Wigner formalism, oscillatory structures represent interference patterns generated by nonclassical correlations encoded in the quasiprobability distribution. The quantity $\mu^{-1}$ therefore defines a characteristic correlation length in phase space, whereas larger values of $\mu$ correspond to increasingly pronounced quantum interference effects.
From the viewpoint of deformation quantization, the oscillatory sector of the Wigner function encodes residual nonclassical correlations that survive the semiclassical limit. Consequently, $\mu^{-1}$ defines an effective correlation length in phase space, while $\mu$ measures the strength of quantum interference effects contributing to the effective cosmological dynamics.
It controls the characteristic modulation scale of the smeared phase-space quantum state, which may emerge from semiclassical WDW wave packets, squeezed cosmological states, or coherent superpositions of gravitational degrees of freedom, and for which oscillatory phase-space patterns arise in the corresponding Wigner representation.

\section{Physical observables} 

One can now turn the attention to Eq.~\eqref{cmbh} as to get a more precise interpretation of the quantum effects from the sinusoidal modulation (cf. Eq.~\eqref{imWBmmws4QC}) on the computation of $h$. By turning Eq.~\eqref{cmbh} into an iteration equation for obtaining $h_{_{LS}} \to \tilde{h}(\mu,\sigma)$, one would have
\footnotesize
\begin{eqnarray}
 h_{_{LS}}\to\tilde{h}(\mu,\sigma)
 &=& \frac{ \int_{z_{_{\rm LS}}}^\infty {dz} {\left[\left(\omega_m\,b_{\sigma-1}(\mu)\, (1+z)^3 + \omega_r\,b_{\sigma}(\mu)\,(1+z)^4\right) \left(1 + {3\omega_b}/({4\omega_\gamma}{(1+z)})\right) \right]^{-1/2}} }{\sqrt{3} \theta_s { \int_{0}^{z_{_{\rm LS}}}
 dz\, \left[\omega_m\,b_{\sigma}(\mu)\,(1+z)^3+(h_{_{LT}}^2-\omega_m)\,\right]^{-1/2}}}h_{_{LT}},\quad\label{cmbh2}
\end{eqnarray}\normalsize
again with $b_\kappa(\mu)$ from Eq.~\eqref{imWBmmws5}, and with $w = -1$, $\omega_b=0.0224\pm0.0001$ and $z_{_{\rm LS}}\simeq 1080$. In particular, the sound propagation and late time measurements are assumed to not be affected by quantum corrections.

\begin{figure}[h!]
\includegraphics[scale=.5]{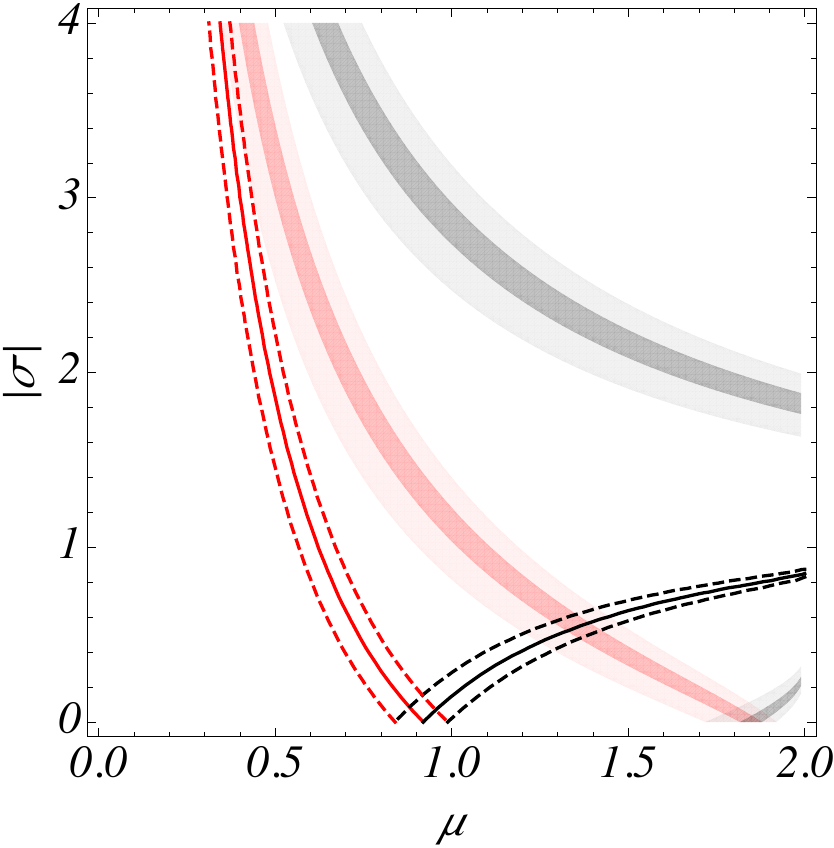}
\renewcommand{\baselinestretch}{1}
\caption{\footnotesize(Color online) First plot depict $\sigma$ and $\mu$ constraints such that $h \sim 0.732 \pm 0.013$ from Eq.~\eqref{cmbh2}. Solid lines are for $h \sim 0.732$, and dashed lines are for upper and lower limits ($\pm 0.013$). Results are confronted with the predictions described in the previous section (less restrictive approach), where black ($\sigma \leq 0$) and red ($\sigma \geq 0$) dark regions show the limits when experimental errors for early time measurements, i.e. $\tilde{h}= 0.673 \pm 0.006$, are included. Correspondently, light regions show the limits when experimental errors for both early and late time measurements, $h_{_{LS}} = 0.673 \pm 0.006$ and $h_{_{LT}} = 0.732 \pm 0.013$, are included.}\label{quarta}
\end{figure}

Fig.~\ref{quarta} shows $\sigma$ and $\mu$ constraints such that $h_{_{LT}} \sim 0.732 \pm 0.013$ at Eq.~\eqref{cmbh2}. The same arguments for the constraint between $\mu$ and $\sigma$ are thus identified through this approach, and confronted with the constraints from the previous one, when it is driven by the experimental inputs. Red and black tones are for positive and negative values of $\sigma$, respectively. Dark black (red) region shows the limits when experimental errors for early time measurements, i.e. $h_{_{LS}} = 0.673 \pm 0.006$, are included. Light black (red) region shows the limits when experimental errors for both early and late time measurements, $h_{_{LS}} = 0.673 \pm 0.006$ and $h_{_{LT}} = 0.732 \pm 0.013$, are included. Computation from Eq.~\eqref{cmbh2} follows from the addressed first approach at Ref.~\cite{Lesgougues} and not only suggests more consistent values for $h$, but also provides the systematic procedure for including a curvature background in the analysis. As depicted in Fig.~\ref{quarta}, results are confronted with predictions from the less restrictive (second one) approach suggested at Ref.~\cite{Lesgougues}.

From the first plot of Fig.~\ref{quarta}, one notices a weak convergence between both approaches just for higher values of the lapse function parameter ($\sigma > 4$) and for the quantum envelop parameter $\mu \sim 1/2$.
In fact, both approaches are independent: they intend to fit the phenomenological predictions for $h$ at early and late times by different systematic perspectives, as explained in Refs.~\cite{Lesgougues}.
Giving that the missing agreement between such two approaches have been strongly marked in the recent literature \cite{Lesgougues,Hubble1,Hubble2,Hubble3}, additional steps are necessary.

For instance, an artificial remodulation of the quantum envelop could optimize the second approach in order to overcome the soft missing agreement between both of them.
The quantum envelop modulation parameter, $\mu$, should be expected with some value around the unity, $\mu \lesssim 1$, which physically suggest that the considered quantum envelop, $\sim \sin(\mu k x) \sim \sin (p q /\hbar)$ maps the ground state of a cosmological (squared) quantum well. For both approaches, i.e. those ones from Eq.~\eqref{cmbh2} and from the predictions depicted in Fig.~\ref{terca}, obtaining fitting parameters $\mu \gg 1$ would ruins our proposal.

\subsection{Inclusion of curvature}

Within the FLRW framework, spatial curvature enters the Friedmann equation as an additional contribution scaling as $a^{-2}$, thereby affecting both the expansion history and the angular diameter distance to last scattering. This introduces well-known parameter degeneracies in CMB analyses, particularly among $\Omega_{_{K}}$, $H_0$, and the matter density parameters, which initially motivated studies of non-flat cosmologies as a possible avenue to relax the tension between early- and late-time determinations of the Hubble constant \cite{Aghanim,Riess:2020fzl,Arjona2021,Vagnozzi2021}.
However, while non-zero curvature can locally shift parameter posteriors and partially broaden confidence regions, combined analyses with BAO and other late-time geometric probes strongly constrain departures from flatness. In particular, these datasets tend to break the geometric degeneracies present in CMB-only fits, driving $\Omega_{_{K}}$ toward values consistent with spatial flatness and leaving the Hubble tension essentially unresolved \cite{Moresco2022,Hubble1}.
More recently, the role of curvature has therefore shifted from being considered a standalone resolution mechanism to being treated as part of broader model extensions, where it is combined with modified dark-energy sectors or early-Universe physics. In this context, curvature remains an important consistency parameter of the cosmological model, but current observational evidence indicates that it is unlikely to provide a complete explanation of the Hubble tension on its own \cite{Handley2020,Efstathiou2020}.

Curvature contributions can also be incorporated into our analysis. The position of an astronomical source at cosmological distances is specified by its observed redshift $z$ and its angular coordinates on the sky, namely right ascension $\alpha$ and declination $\delta$. In order to identify voids in the three-dimensional distribution of tracers, it is necessary to first transform these observables into comoving Cartesian coordinates $\mathbf{x}$,
\begin{equation}
\mathbf{x}(z,\alpha,\delta)
= (1+z)\,D_A(z)
\begin{pmatrix}
\cos\alpha \cos\delta \\
\sin\alpha \cos\delta \\
\sin\delta
\end{pmatrix},
\tag{103}
\end{equation}
where $D_A(z)$ denotes the comoving angular-diameter distance to a tracer (or to the LS surface) at redshift $z$. This quantity depends on the cosmic expansion history through the Hubble parameter $H(z)$ and on spatial curvature via the parameter $\Omega_{_{K}}$, as conveniently defined in \cite{Huterer},
\begin{equation}\label{eq:angdiamdistance}
 D_A(z_{_{\rm LS}}) = \frac{c}{H_0\,\sqrt{\vert\Omega_{_{K}}\vert}}\,s_K\left(\int_{0}^{z_{_{\rm LS}}} dz\frac{\sqrt{\vert\Omega_{_{K}}\vert}}{H(z)/H_0}\right)~,
\end{equation} with $s_K(x) = x$ in a flat universe, $s_K(x)= \sin(x)$ in a closed universe ($\Omega_{_{K}} < 0$), and finally $s_K(x)= \sinh(x)$ in an open universe ($\Omega_{_{K}} > 0$).
In this case, the constraint Eq.~\eqref{cmbh} becomes
\small \begin{equation}
\tilde{h}(\mu,\sigma) = \frac{\sqrt{\omega_K} \int_{z_{_{\rm LS}}}^\infty {dz} {\left[\left(\omega_m \,b_{\sigma}(\mu)\,(1+z)^3 + \omega_r\,b_{\sigma-1}(\mu)\,(1+z)^4\right) \left(1 + {3\omega_b}/({4\omega_\gamma}{(1+z)})\right) \right]^{-1/2}} }{\sqrt{3} \theta_s s_K\left[ \sqrt{\omega_K}{ \int_{0}^{z_{_{\rm LS}}}
 dz\, \left[\omega_m\,b_{\sigma}(\mu)\,(1+z)^3+(h^2-\omega_m)(1+z)^{-3(1+w)} \right]^{-1/2}}\right]}h_{_{LT}}\label{cmbhcur}
\end{equation}\normalsize
with $\omega_K =\Omega_{_{K}} h^2$.

\begin{figure}[h!]
\includegraphics[scale=.6]{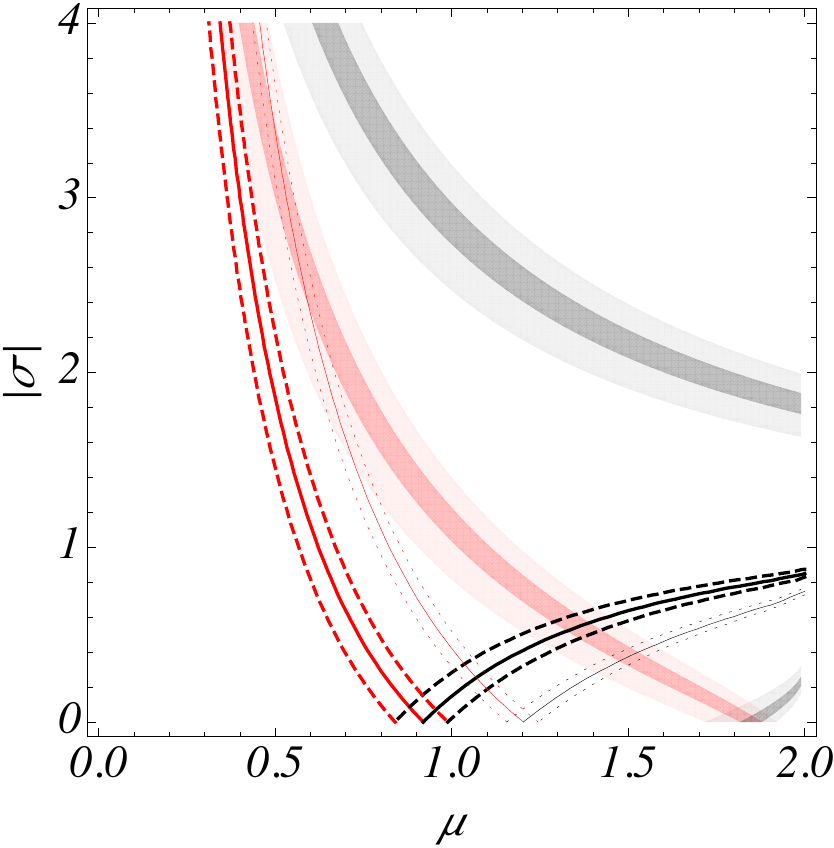}
\renewcommand{\baselinestretch}{1}
\caption{\footnotesize(Color online) $\sigma$ and $\mu$ constraints considering the inclusion of curvature parameterized by $\vert\Omega_{_{K}}\vert < 0.005$ at Eq.~\eqref{cmbhcur} (region bounded by thin lines). Results are confronted with those depicted from Fig.~\ref{quarta}, for $\Omega_{_{K}} = 0$ (thick lines). Solid lines are for $h_{_{LT}} \sim 0.732$, and dashed lines are for upper and lower limits ($\pm 0.013$). }\label{quinta}
\end{figure}

The $\sigma$ and $\mu$ constraints for a closed universe parameterized by $-0.005 < \Omega_{_{K}} < 0$, as given in Eq.\eqref{cmbhcur}, are shown in Fig.\ref{quinta}, where limits for $\Omega_{_{K}} = -0.005$ as depicted by thin lines, push the bounds to the right. In this case, no additional modulation of the quantum envelope parameter would required. Essentially, $\Omega_{_{K}}$ highly contributes to the tuning of the quantum related parameters, $\mu$ and $\sigma$, as it does not alter the fundamental aspects of our quantum proposal.

Now, another argument which favors our analysis can be set for the value of the lapse parameter $\sigma$. Results from Fig.~\ref{quinta} suggest that $\sigma$ could be set equal to zero (or even any integer value) from the beginning of our analysis, showing that the gauge of the quantum potential is not determinant for the consistence with the phenomenological result, as depicted by the appealing constraint between $\sigma$ and $\mu$ from Fig.~\ref{quinta}.

\begin{figure}[h!]
\includegraphics[scale=.7]{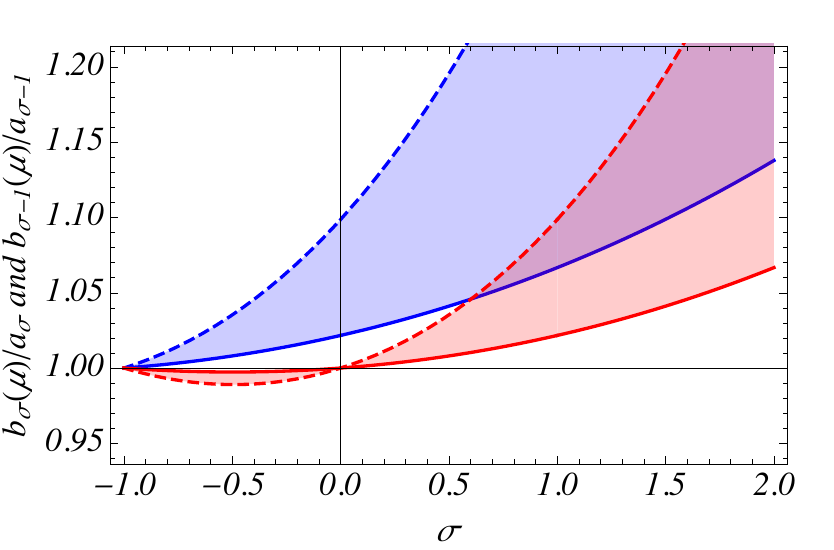}
\renewcommand{\baselinestretch}{1}
\caption{\footnotesize(Color online) Fluctuations of $b_{\sigma}(\mu)$ (blue region) and $b_{\sigma-1}(\mu)$ (red region) (over $a_\sigma$ and $a_{\sigma-1}$, for matter and radiation, respectively) as function of $\sigma$, in the region of $\mu$ between $1/2$ (solid lines) and $1$ (dashed lines). Results are in correspondence with the second plot from Fig.~\ref{quarta}.}\label{final}
\end{figure}

Of course, and still more relevant, that Planck data for calibrating $\Lambda$CDM and classical cosmology outputs from that should also not be significantly affected in the context of our approach, otherwise, again, our proposal would not make sense.
Coefficients $b_{\sigma}(\mu)$ (matter) and $b_{\sigma-1}(\mu)$ (radiation) from Eq.~\eqref{imWBmmws5} play the essential role for such an expected consistency of our analysis. Fig.~\ref{final}, in correspondence with the second plot from Fig.~\ref{quarta}, show the fluctuations of $b_{\sigma}(\mu)$ (blue region) and $b_{\sigma-1}(\mu)$ (red region) (over $a_\sigma$ and $a_{\sigma-1}$, respectively) as function of $\sigma$, in the region of $\mu$ between $1/2$ (solid lines) and $1$ (dashed lines).
Considering the correspondence of $a_\sigma$ and $a_{\sigma-1}$ with matter and radiation energy density outputs from Planck measurements, the maximal magnitude of fluctuations (for $\sigma < 1$ and $\mu=1/2$) identified in the range covered by the parameters from Fig.~\ref{final}, from $0\%$ to $\sim 6.5\%$ for matter, and from $0\%$ to $\sim 2.0\%$ for radiation, are acceptably covered by the Planck data statistical deviations.

Although the numerical value of the correction depends quantitatively on the parameters $\mu$, $\sigma$, and $\Omega_{_{K}}$, the mechanism itself does not originate from a singular fine-tuned configuration. The effective modification of the cosmological dynamics emerges from the existence of residual phase-space quantum corrections. The present analysis should therefore be interpreted as a proof-of-principle demonstration of the mechanism, while a complete statistical exploration of admissible regions in the $(\mu,\sigma,\Omega_{_{K}})$ parameter space is left for next steps, within which the resulting quantum corrections remain phenomenologically relevant.
Our conclusion is that the appropriate tuning of the wave function parameter, $\mu$, which can be re-scaled in the quantum Wigner function $\propto \mbox{Sn}(\mu ,k,x)$, and the inclusion of tiny contributions of curvature, both reduces the Hubble tension without disrupting the consistency with phenomenological data from both early- and late-time observations.

\section{Conclusions}

\hspace{1 em}A theoretical resolution for the so-called Hubble tension problem \cite{cosmoverse} has been investigated within the framework of Weyl-Wigner quantum mechanics. Essentially, our analysis has shown that quantum effects can suppress the Hubble tension divergence between early- and late-time cosmological predictions.
Considering the prevalence of the standard $\Lambda$CDM model, our results suggest that the framework here considered can be extended to more general quantum cosmological scenarios, including those that incorporate spatial curvature and modifications to the dark sector. 

This includes, for instance, a compelling candidate for a renormalizable quantum gravity theory (at high energies) provided by Ho\v{r}ava-Lifshitz (HL) gravity \cite{Hora09,Horava:2010zj,Blas:2009qj,Blas:2010hb,Soti2009gy,Soti2009bx}.
HL gravity \cite{Mukohyama:2010xz,Saridakis:2011pk,BerBer,Mukohyama:2009mz,Soti2009bx,Maeda:2010ke,Wang:2009rw,Wang:2009rw2}
engenders a framework through which the notion of time \cite{Page83,Vilenkin94,VilenBerto1,VilenBerto2} and the meaning of quantum probabilities \cite{Page83,Vilenkin94,Ander12,Merali13}, the initial singularity problem and its ensued conditions for the onset of inflation \cite{Linde84,Rubakov84,Vilen84,Bousso96,Linde98,Hartle83,Grishchuk}, have been investigated through a series of cosmological models \cite{Mukohyama:2010xz,Saridakis:2011pk,Mukohyama:2009mz,Soti2009bx,Wang:2009rw,Wang:2009rw2} which include the Hubble tension issues \cite{Berechya20}.

In spite of being driven by phase-space quantum mechanics, our results can be embraced by an open landscape of quantum cosmological scenarios. In the absence of a decoherence mechanism, the quantum-to-classical transition in our framework is tightly constrained by the choice of the quantum cosmological model --- particularly by the analytical tractability of the WDW wave functions and the associated Wigner functions. Notably, the mathematical manipulability of the Weyl transform of sinusoidal test Wigner functions enables a broad, model-independent exploration of quantum effects.

For classical potentials of the form $\mathcal{V}(x) = \sum_{\kappa} a_\kappa x^{-\kappa}$ --- which includes a wide class of both classical and quantum integrable systems --- a multiple rescaling of the coordinate $x$ driven by the contributions from each $\kappa$ term appears essential to generating modified {\em plateaux} patterns, as expected in the evolution of the Hubble parameter from early to late cosmological times. Within this context, quantum-origin corrections to the Einstein-Friedmann equation, mediated by generalized localized phase-space quantum states, predict a smooth transition in the Hubble parameter $H_0$ across cosmological epochs, thereby preserving compatibility with current phenomenological constraints.

\emph{Acknowledgments -- The author acknowledges support from the Brazilian agency CNPq (Grant No. 301485/2022-4, National Council for Scientific and Technological Development -- CNPq). The author also thanks the Departamento de F\'isica e Astronomia, Faculdade de Ci\^{e}ncias da Universidade do Porto, for its hospitality, and Prof. Orfeu Bertolami for stimulating scientific discussions.}

\end{document}